\documentclass{article}
\usepackage{spconf,amsmath,graphicx}
\usepackage{epsfig}

\def\(({\left(}
\def\)){\right)}
\def\[[{\left[}
\def\]]{\right]}

\newcommand{\be}{\begin{equation}}
\newcommand{\ee}{\end{equation}}
\newcommand{\bea}{\begin{eqnarray}}
\newcommand{\eea}{\end{eqnarray}}

\newcommand{\bX}{{\textbf {X}}}

\newcommand{\bx}{{\textbf {x}}}

\newcommand{\bs}{{\textbf {s}}}
\newcommand{\by}{{\textbf {y}}}

\title{Compressed Sensing under Matrix
  Uncertainty: \\ Optimum Thresholds and Robust
  Approximate Message
  Passing} \name{Florent Krzakala$^{1}$\sthanks{The research
    leading to these results has received funding from the European
    Research Council under the European Union's $7^{th}$ Framework
    Programme (FP/2007-2013)/ERC Grant Agreement 307087-SPARCS.}, Marc
  M\'ezard~$^{2}$ and Lenka Zdeborov\'a~$^{3}$} \address{ $^{1}$ CNRS
  and ESPCI ParisTech 10 rue Vauquelin, UMR 7083 Gulliver Paris 75005
  France \\ $^{2}$ Ecole Normale Sup\'erieure, 45 rue d'Ulm, Paris,
  France, and LPTMS-CNRS, Univ. Paris Sud \\
  $^{3}$ IPhT, CEA Saclay, and URA 2306 CNRS, 91191 Gif-sur-Yvette,
  France}
\begin{document}
%
\maketitle
\begin{abstract}
  In compressed sensing one measures sparse signals directly in a
  compressed form via a linear transform and then reconstructs the
  original signal. However, it is often the case that the linear
  transform itself is known only approximately, a situation called
  “matrix uncertainty”, and that the measurement process is
  noisy. Here we present two contributions to this problem: first, we
  use the replica method to determine the mean-squared error of the
  Bayes-optimal reconstruction of sparse signals under matrix
  uncertainty. Second, we consider a robust variant of the approximate
  message passing algorithm and demonstrate numerically that in the
  limit of large systems, this algorithm matches the optimal
  performance in a large region of parameters.
\end{abstract}
\section{Introduction}
Compressed sensing (CS) is designed to measure sparse signals directly
in a compressed form by acquiring a small (with respect to the
dimension of the signal) number of random linear projections of the
signal, and subsequently reconstructing computationally the signal. In
particular, it has been shown \cite{CandesTao:06,Donoho:06} that this
reconstruction is possible and computationally feasible in many cases
using $\ell_1$-minimization based algorithms. It is often the case in
practical applications that the linear transform itself is known only
approximately; for instance this is the case if calibration has not been
done perfectly. Several works have analyzed the performance of
$\ell_1$-regularization based algorithm under matrix uncertainty,
e.g. \cite{ZhuLeus11,RosenTsy10} and references therein.  In this
paper we shall use instead a Bayesian approach. In this
framework CS under matrix uncertainty was studied recently by
\cite{SCHNITER}. In particular the authors of \cite{SCHNITER} propose
a generalization of the approximate message passing algorithm (AMP)
\cite{DonohoMaleki09,Rangan10b,KrzakalaPRX2012} that treats matrix
uncertainty (MU-AMP) and outperforms other methods. Empirically, they show that the
MU-AMP algorithm performs near oracle bounds.

Our goal in this paper is twofold. First we compute the best
theoretically possible performance of reconstruction algorithms under
matrix uncertainty using the replica
method~\cite{MezardParisi87b,RanganFletcherGoyal09,GuoBaron09,KabashimaWadayama09,KrzakalaPRX2012,KrzakalaMezard12}, and
show that in a large region of parameters this performance is indeed
matched by MU-AMP. And second we consider a slightly different
algorithm that we refer to as robust-AMP algorithm (using a minimal
change first suggested in \cite{DonohoMaleki09}) and show that it is
asymptotically equivalent to MU-AMP and thus also it reaches Bayes
optimal MSE while not assuming the knowledge of the measurement noise,
nor the level of matrix uncertainty.

\section{Definitions}
\label{sec:Definition}
In our analysis we consider a sparse $N$-dimensional signal $\bx$ having on
average $K$ non-zero iid components chosen from a distribution $\phi(x)$. Defining the
density $\rho=K/N$ we have
\be P(\bx) = \prod_{i=1}^N P(x_i) =\prod_{i=1}^N [
\rho\phi(x_i) + (1-\rho) \delta(x_i) ] \, .\label{Px} \ee
Note that our analysis applies as well to non-iid signals with
empirical distribution of components converging to $\phi(x)$, as
discussed e.g. in~\cite{DonohoJavanmard11}. For concreteness, here we
shall use a Gaussian $\phi(x)$ of zero mean and unit
variance. Although our results depend quantitatively on the form of
$\phi(x)$, the overall qualitative picture is robust with respect to
this choice. We further assume that the parameters of $P(\bx)$ are
known (if this is not the case they can be learned via
expectation-maximization~\cite{KrzakalaPRX2012,KrzakalaMezard12,VilaSchniter11}). One
could use an approximately sparse signal as in
\cite{BarbierKrzakalaAllerton2012} as well.

The $M$
measurements $y_\mu$ are obtained via noisy linear random projections of
the signal
\be y_\mu = \sum_{i=1}^N F^0_{\mu i }s_i + \xi_{\mu}   \, , \quad
\quad \mu=1,\dots,M  \, .
\label{def} \ee 
where we model the noise $\xi_{\mu}$ as a Gaussian random variable
with mean $0$ and variance $\Delta$, and where ${\bf F^0}$ has iid random
components of mean $0$ and variance $1/N$. The parameter $\alpha=M/N$ is the
measurement, or sampling, rate.

Additionally, we consider that the matrix ${\bf F^0}$ is not known
perfectly, and that we know only its noisy version:
\be F'_{\mu i} = \frac{{F_{\mu i}^0+X_{\mu
      i}\sqrt{\eta}}}{\sqrt{1+\eta}}\, ,
\ee 
where $X_{\mu i}$ is a white noise with mean $0$ and variance $1/N$, and where $\eta$ measures the uncertainty on the matrix. When
$\eta\to 0$, the measurement matrix is perfectly known (this is the
usual compressed sensing situation) while for $\eta \to \infty$
nothing is known about the measurement matrix. Assuming a Gaussian
distribution for the matrix elements and the noise $\bX$, we obtain
the posterior probability of one matrix element $P(F_{\mu i}^0|F_{\mu
i}')$ to be a Gaussian with mean and variance
\be
F_{\mu i} = \frac{F'_{\mu i}}{\sqrt{1+\eta}}\, , \quad \quad S_{\mu i} = \frac{\eta}{N(1+\eta)}\, . 
\ee

Our goal is to reconstruct the signal $\bs$ based on the knowledge of $\by$ and
${\bf F'}$. The Bayes-optimal estimate of a signal $\bx^{\star}$, that
minimizes the mean-squared
error (MSE) ${\rm MSE}=\sum_{i=1}^N (s_i - x_i^{\star})^2 /N$ with the
original signal $\bs$, is given by
\be x^{\star}_i=\int \text{d} x_i \,x_i\, \nu_i(x_i) \,
,\label{average_marginal} \ee
where $\nu_i(x_i)$ is the marginal distribution of $x_i$
with respect to:
\be P(\bx | \by, {\bf F'}) = \frac{P(\bx)}{Z} \int {\rm d}{\bf F^0}
P({\bf F^0}|{\bf F'}) 
e^{-\frac{1}{2\Delta}(\by - {\bf F^0} \bx )^2}\, .\label{p_bayes} \ee

\section{Replica method: Optimum reconstruction bounds}

The MSE obtained from the Bayes-optimal approach can be computed in
the limit of large $N$ via the replica method. Although this method is
in general
non-rigorous, it is sometimes possibles to prove that the results
derived from it are exact, see~\cite{MezardParisi87b,RanganFletcherGoyal09,GuoBaron09,KabashimaWadayama09,KrzakalaPRX2012,KrzakalaMezard12}. 
We shall leave out a detailed derivation and refer instead
to~\cite{KrzakalaPRX2012,KrzakalaMezard12} for a very similar
computation. The result is that in the large signal limit, the MSE is given by
\be
{\rm MSE}(\alpha,\rho,\Delta,\eta)=\arg\!\max_{E} \Phi(E) \, ,
\ee
where the ``potential'' $\Phi(E)$ is given by
\bea 
\Phi(E)\!\!\!\!\!\! &=&\!\!\!\!\!\! 
-\frac{\alpha}{2} \!\!\left[ \log{[\Delta \!+\! E \!+\! (\rho-E) D]} \!\! +\!\!
  \frac{\Delta+\rho}{[\Delta \!+\! E \!+\! (\rho-E) D]}\right] \nonumber \\ &+& (1-\rho)
\int {\cal D}z \, \log{\left[ 1 -\rho + \frac{\rho}{\sqrt{ m
        +1}} e^{\frac{z^2 m}{2( m +1)}} \right]}\nonumber \\ &+&
\rho \int {\cal D}z \, \log{\left[ 1 -\rho + \frac{\rho}{\sqrt{
        m +1}} e^{\frac{z^2 m}{2}} \right]} \, , 
\label{replica}
\eea 
with 
\be
D= \frac{\eta}{1 +\eta}\, , \quad \quad  m=\frac{\alpha (1-D)}{ \Delta +
  E + (\rho-E) D}\, .
\ee
Note how closely related is this expression to the one derived in
\cite{KrzakalaMezard12} for the usual compressed sensing
with known measurement matrix (which is recovered taking $\eta=0$). 

\begin{figure}[!t]
  \begin{center}
    \hspace{0.4cm}\includegraphics[width=0.95\linewidth]{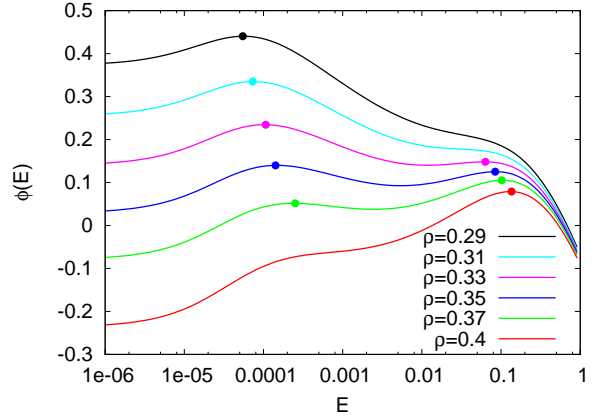}
    \caption{The potential $\Phi(E)$, eq.~(\ref{replica}), for
      measurement rate $\alpha=0.5$, matrix uncertainty $\eta=10^{-4}$, and
      noise $\Delta=10^{-10}$, for different values
      of density $\rho$. The global maximum $\Phi(E^*)$ determines
      the values of the Bayes-optimal ${\rm MSE}=E^*$. 
      As the density varies the value $E^*(\rho)$ has a discontinuity
      which is usually referred to as a ``first-order'' or
      ``discontinuous'' phase transition.  
      AMP algorithms are performing a steepest ascent of the potential and
      get trapped at the maximum with the largest value of E,  instead
      of reaching the global
      one (see e.g. the case $\rho=0.33$). The appearance of this
      trapping local maxima is referred to as the ``spinodal'' transition.}
    \label{fig:potential}
  \end{center}
\end{figure}
\begin{figure}[!t]
  \begin{center}
    \includegraphics[width=1.01\linewidth]{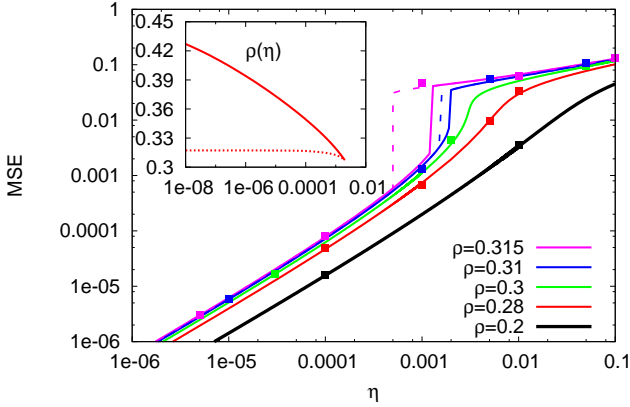}
    \caption{Bayes-optimal MSE for $\alpha=0.5,\, \Delta=10^{-10}$ as a
      function of matrix uncertainty $\eta$ for different densities
      $\rho$. The MSE of the trapping local maximum is shown in dashed line. The results of
      the AMP algorithm on some instances are also shown (points, done
      with $N=20000$). The results of these AMP runs (points) agree
      perfectly with the prediction (lines). In the inset, we show the location of
      the two phase transitions in the $\rho,\, \eta$
      plane: the upper line determines the ``first-order'' threshold
      beyond which the optimal-MSE suddenly degrades, while the lower line is
      the ``spinodal'' transition that marks the degradation of the
      performance of AMP. Note that the spinodal is
      almost independent of the noise $\eta$ in a large range of
      values.}
    \label{fig:MSE}
  \end{center}
\end{figure}

\begin{figure}[!t]
  \begin{center}
      \includegraphics[width=1\linewidth]{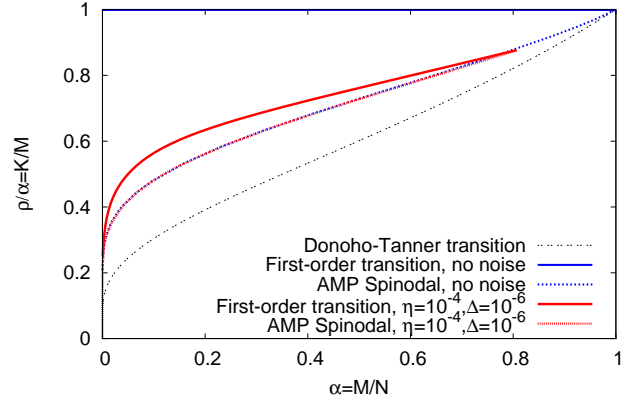}
      \caption{The phase diagram ``a la'' Donoho-Tanner
        \cite{Donoho05072005} in the limit of large signals, $N\to \infty$. The
        horizontal axis is the sampling ratio $M/N$ and the vertical
        one measures the gain $K/M$ provided by compressed sensing.
        The dotted black line is the transition for the 
        $\ell_1$ reconstruction \cite{Donoho05072005} in zero noise.
        The blue lines (from \cite{KrzakalaPRX2012,KrzakalaMezard12})
        show the location of the spinodal (dotted blue) and first-order
        (full blue) phase transitions for the noiseless
        Bayesian approach. While a perfect reconstruction (MSE=$0$)
        can be obtained in principle until $\rho=\alpha$, the 
        AMP algorithm allows such perfect reconstruction only up to the
        spinodal line (dotted blue). The noisy counterpart
        of these two transitions is shown in red, using
        $\Delta=10^{-4}$ and $\eta=10^{-6}$. While the region where a
        good reconstruction is possible (below the full red line) has
        shrunk a lot, the spinodal line (dotted red line) that marks
        the onset of good performance of
        AMP is left almost unchanged below
        $\rho \approx 0.8$ demonstrating the robustness of the
        AMP  to noises.}
    \label{fig:phase}
  \end{center}
\end{figure}

The potential (\ref{replica}) is shown in Fig.~\ref{fig:potential}.
Maximizing the potential, one obtains the Bayes-optimal MSE which we
show in Fig.~\ref{fig:MSE} as a function of $\eta$ for different
densities $\rho$. We see a phenomenology similar to that described in
~\cite{KrzakalaMezard12,BarbierKrzakalaAllerton2012}
for other types of noise. Because of a two-maxima shape of the
function $\Phi(E)$, the Bayes-optimal MSE displays a sharp transition
separating a region of parameters 
with a small MSE, comparable to $\max(\Delta,\eta)$, from a region with
a large $O(1)$ value of the MSE. In this second region of parameters compressed
sensing technics --regardless of the reconstruction method--
will not be useful.  This is a ``first order'' phase transition. 

Another important transition (the ``spinodal'') that can be studied from the form of the
potential function $\Phi(E)$ is the appearance of a high-MSE local
maxima. Since the AMP algorithm performs a steepest ascent in the
$\Phi(E)$ function (see e.g. \cite{KrzakalaMezard12}) this transition separates the region of parameters in
which the performance of AMP will be asymptotically optimal from the
one where AMP is suboptimal. An important feature of this transition,
shown in Fig.~\ref{fig:phase}, is that its position is only very weakly dependent on the value
of the matrix uncertainty. This means that the performance of AMP is
extremely robust to matrix uncertainty $\eta$. The same result was also obtained for
additive measurement noise in~\cite{KrzakalaMezard12} and for approximate sparsity
in~\cite{BarbierKrzakalaAllerton2012}.

\section{Robust Message-Passing algorithm}
The Bayesian approach to compressed sensing combined with belief
propagation based reconstruction algorithm leads to the so-called
approximate message passing (AMP) algorithm as first derived in
\cite{DonohoMaleki09} for the minimization of $\ell_1$, and
subsequently generalized in
\cite{DonohoMaleki10,Rangan10b,KrzakalaMezard12}.  This approach was
adapted by Parker, Cevher and Schniter to treat the matrix uncertainty
MU-AMP in \cite{SCHNITER}. Here we shall consider a version of the
canonical AMP, that we call robust-AMP, that turns out to be robust to
noisy measurement and matrix uncertainty, so that it can be used
indifferently of the presence or absence of noise and matrix
uncertainty. We show in the next section that robust-AMP and MU-AMP
are equivalent in the limit of infinite systems.

For every measurement component $\mu$ we define one real number 
$\omega_\mu$, for each signal component $i$ we define four real
numbers $\Sigma_i$, $R_i$,
$a_i$, $v_i$. These quantities are updated as follows (for the
derivation with these notations see \cite{KrzakalaMezard12}):
\bea
 V^{t+1}  &=& \frac{1}{M} \sum_{\mu} (y_{\mu} - \omega^{t}_\mu)^2  \, , \label{TAP_ga} \\
\omega^{t+1}_\mu  &=& \sum_i F_{\mu i} \, a^t_i -\frac{  (y_\mu-\omega^t_\mu)}{V^t} \sum_i F_{\mu i}^2\,v^t_i \, , \label{TAP_al}  \\
   (\Sigma^{t+1}_i)^2&=&\frac{V^{t+1}} { \sum_\mu {F^2_{\mu i}} }\, ,
   \label{TAP_U}\\
   R^{t+1}_i&=& a^t_i + \frac{\sum_\mu F_{\mu i} (y_\mu - \omega^{t+1}_\mu)}{ \sum_\mu F_{\mu i}^2}\, , \label{TAP_V}\\
   a^{t+1}_i &=&   f_a\left((\Sigma^{t+1}_i)^2,R^{t+1}_i\right) ,  \label{TAP_a}\\
   v^{t+1}_i &=& f_c\left((\Sigma^{t+1}_i)^2,R^{t+1}_i\right) \,
   . \label{TAP_v} 
\eea 
Where only the functions $f_a$ and $f_c$ depend explicitly on the
model $P(x)$: 
\bea
\!  f_a(\Sigma^2,R)  \!\! \!\! &\equiv& \!\!\!\!\int {\rm d}x\,  x \,  {\cal M}(\Sigma^2,R,x)\, , \label{f_a_gen}\\
\! f_c(\Sigma^2,R)  \!\!\!\!&\equiv& \!\!\!\!\int {\rm d}x\, x^2 \, {\cal
  M}(\Sigma^2,R,x) - f^2_a(\Sigma^2,\!R)\, , \label{f_c_gen} 
\eea
where ${\cal M}(\Sigma^2,R,x)$ is the following probability distribution
\be
         {\cal M}(\Sigma^2,R,x)  \equiv \frac{1}{\hat Z(\Sigma^2,R)} P(x) \frac{1}{\sqrt{2\pi} \Sigma} e^{-\frac{(x-R)^2}{2\Sigma^2}}.
\ee
The explicit expression for $f_a$ and $f_c$ for the Gauss-Bernoulli signal is given in
\cite{KrzakalaMezard12} while the case of approximate sparsity was
considered in \cite{BarbierKrzakalaAllerton2012}. The above equations
are initialized with $a_i^{t=0}=0$, $v_i^{t=0}=\rho$,
$\omega_\mu^{t=0}=y_\mu$, then the equations
are iterated till convergence.

The difference between the present algorithm with respect to the more
common version of AMP is in the way the estimate of the current error
on the measurement element is computed in eq.~(\ref{TAP_ga}).  Most
previous works (e.g. \cite{Rangan10b,KrzakalaMezard12}) used a
$\mu$-dependent vector $V_{\mu}^{t+1}=\Delta+\sum_i F_{\mu i}^{2}
v^{t}_i$ in the case of noisy measurement with a perfectly known
matrix, whereas the original paper \cite{DonohoMaleki09} used a value
precomputed by the state evolution. The modification done by the
authors of ref.~\cite{SCHNITER} to incorporate the effect of matrix
uncertainty is (in our notation) $V_{\mu}^{t+1}=\Delta+\sum_i F_{\mu
  i}^{2} v^{t}_i +\sum_i (b_i+a_i^2) \eta /[N(1+\eta)]$. As shown in
ref.~\cite{SCHNITER} this leads to a very efficient algorithm with
matrix uncertainty. The expression we use instead in
eq.~(\ref{TAP_ga}) was also proposed in \cite{DonohoMaleki09}. Perhaps
surprisingly, it is equivalent to the one of ref.~\cite{SCHNITER} when
the system size $N\to \infty$ even in the case of matrix
uncertainty. The advantage of expression~(\ref{TAP_ga}) is that it is 
not need to explicitly know the value of $\eta$ and
$\Delta$\footnote{Of course $\Delta$ and $\eta$ can be learned with
  expectation maximization within the MU-AMP
  \cite{KrzakalaPRX2012,KrzakalaMezard12,VilaSchniter11}, but this
  adds considerable computational time.}: the algorithm automatically
incorporates the errors coming from the uncertainty on the matrix and
the measurement; and hence we are using the same code regardless the
presence or absence of noise and/or matrix uncertainty.
We thus refer to algorithm (\ref{TAP_ga})-(\ref{TAP_v}) as the ``robust''-AMP.

\section{Density Evolution}
\label{sec:DensityEvolution}
\begin{figure}[!t]
  \begin{center}
      \includegraphics[width=1\linewidth]{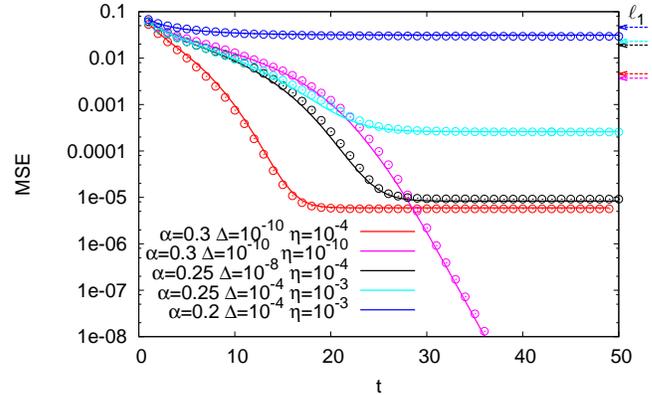}
      \caption{Comparison of the time evolution of MSE computed with
        density evolution and the one found numerically using
        robust-AMP for systems of size $N=25000$, $\rho=0.1$.  The
        agreement between theoretical predictions
        (eq.~(\ref{evolution}), full line) with the data from the AMP
        algorithm (points) is algorithm. The arrows show the MSE
        obtained using $\ell_1$ minimization on the same instances,
        for comparison. The better performance of the AMP approach
        over $\ell_1$ is clear.}
    \label{fig:density}
  \end{center}
\end{figure}

The AMP approach is amenable to asymptotic ($N\to \infty$)
analysis in the case of i.i.d. random measurement matrices using a method known as
the ``cavity method'' (in statistical physics)
\cite{MezardParisi87b}, the ``density evolution'' in
coding \cite{RichardsonUrbanke08}, and the ``state evolution'' in the
context of CS \cite{DonohoMaleki09,BayatiMontanari10}. We shall not detail
the computation here, it goes in the same lines as in \cite{KrzakalaMezard12}. Given
the parameters $\rho$, $\alpha$, $\eta$, $\Delta$, the MSE follows:
\be
E^{t+1} = \int {\rm d}x \, P(x) \int  {\cal D} z   f_c(1/ m^t, x +
z /\sqrt{m^t})\, ,\label{evolution} 
\ee
where $m^t$ follows eq.~(\ref{replica}), with $E^{t}$ on the right
hand side, $E^{t=0}=\rho$, ${\cal D} z$ is a Gaussian integral, and
$f_c$ is defined by eq.~(\ref{f_c_gen}). 
The comparison between the evolution of the MSE in the algorithm and
eq.~(\ref{evolution}) is shown on Fig.~\ref{fig:density}; the agreement
is very good. Note that we have also applied the density evolution to
the AMP algorithm with matrix uncertainty of \cite{SCHNITER}, and
found that it obeys asymptotically the same equation
(\ref{evolution}). More specifically, the term $V^{t}$ from eq.~(\ref{TAP_ga}) evolves as
$\Delta + E^t +(\rho-E^t) D$ in the limit $N\to \infty$ both for the expression in ref.~\cite{SCHNITER}
and in eq.~(\ref{TAP_ga}). 
From eq.~(\ref{evolution}), one can also derive that the evolution is
equivalent to a steepest ascent of the potential $\phi(E)$ obtained
from the replica method in eq.~(\ref{replica}).  This underlines the
importance of the spinodal transition illustrated on
Figs.~\ref{fig:potential}, \ref{fig:MSE}, and \ref{fig:phase}.  In
particular we see that the region where the AMP converges to the
Bayes-optimal value of the MSE is quite large, and notably larger than
the region in which the $\ell_1$ minimization is able to give precise
reconstruction. Another point worth noting is that the location of the
spinodal depends only very weakly to the value of the noise, for a
large range of matrix and measurement noise (see inset of
Fig.~\ref{fig:MSE} and Fig.~\ref{fig:phase}): this shows that the
robust from of the AMP algorithm is indeed robust to noise(s).
\section{Conclusions}
\label{sec:Discussions}
We have computed the Bayes-optimal value of the MSE for the
reconstruction of sparse Gauss-Bernoulli signals in presence of matrix
uncertainty with the replica method, and consider a variant of the AMP
algorithm robust to such uncertainty and to measurement noise.
Finally, we have shown that AMP allows one to match the optimum MSE in
a large region of parameters, and that the region is very weakly
sensitive to measurement or matrix noises. Note that the present
analysis applies to random i.i.d.  measurement matrices; it is also
possible to use the seeded spatially coupled measurement matrices of
\cite{KrzakalaPRX2012,KrzakalaMezard12,DonohoJavanmard11} and this
would lead to an even larger region of optimality-matching performance
for AMP.

\vfill\pagebreak

\bibliographystyle{IEEEbib}
\bibliography{refs}

\end{document}